\documentclass[useAMS,usenatbib]{mn2e}
\usepackage{aas_macros,graphicx,times,multirow}
\def\xmm {\emph{XMM-Newton}}

\def\swift {\emph{Swift}}

\def\rxte {\emph{RXTE}}
\def\src {1E\,1547.0--5408}

\def\kes {1E\,1841--045}

\def\rxs {1RXS\,J1708--4009}

\def\xte {XTE\,J1810--197}

\def\wes {CXOU J164710.2--455216}
\def\sgra{SGR\,1806--20}
\def\sgrb{SGR\,1900+14}

\def\sgrd{SGR\,1627--41}

\def\sgrf{SGR\,0501+4516}

\def\int{\emph{INTEGRAL}}

\def\flux {\mbox{erg cm$^{-2}$ s$^{-1}$}}
\def\lum {\mbox{erg s$^{-1}$}}

\def\ss {\mbox{s~s$^{-1}$}}
\def\cm2 {\mbox{cm$^{-2}$}}

\def\rc {\rm}

\title[The bursting/outbursting AXP \src]  {The 2008 October \emph{Swift}
detection of X-ray bursts/outburst from the transient SGR-like AXP \src}
\author[G.~L.~Israel et al.] {G.~L.~Israel,$^{1}$\thanks{E-mail:
gianluca@mporzio.astro.it} P.~Esposito,$^{2,3}$ N.~Rea,$^{4,5}$ S.
Dall'Osso,$^{1}$ F.~Senziani,$^{2,6}$
\newauthor P.~Romano,$^{7}$  V.~Mangano,$^{7}$ D.~G\"otz,$^{8}$  S.~Zane,$^9$ A.~Tiengo,$^2$ D.~M.~Palmer,$^{10}$
\newauthor H.~Krimm,$^{11,12}$ N.~Gehrels,$^{11}$  S.~Mereghetti,$^2$ L.~Stella,$^{1}$ R.~Turolla,$^{13, 9}$
\newauthor S.~Campana,$^{14}$ R.~Perna,$^{15}$ L.~Angelini$^{11}$ and A.~De Luca,$^{2,6}$
\smallskip\\
$^1$INAF/Osservatorio Astronomico di Roma, via Frascati 33, 00040 Monteporzio Catone, Italy\\
$^2$INAF/Istituto di Astrofisica Spaziale e Fisica Cosmica - Milano, via E.~Bassini 15, 20133 Milano, Italy\\
$^3$INFN - Istituto Nazionale di Fisica Nucleare, Sezione di Pavia, via A.~Bassi 6, 27100 Pavia, Italy\\
$^4$Institut de Ciencies de l'Espai (ICE-CSIC, IEEC), Campus UAB, Facultat
de Ciencies, Torre C5-parell, 2a planta, 08193, Barcelona, Spain\\
$^5$University of Amsterdam, Astronomical Institute Anton Pannekoek, Kruislaan 403, 1098~SJ Amsterdam, The Netherlands\\
$^6$IUSS - Istituto Universitario di Studi Superiori, viale Lungo Ticino Sforza 56, 27100 Pavia, Italy\\
$^7$ INAF/Istituto di Astrofisica Spaziale e Fisica Cosmica - Palermo, via Ugo La Malfa 153, 90146 Palermo, Italy\\
$^8$CEA Saclay, DSM/Irfu/Service d'Astrophysique, Orme des Merisiers, B\^at.\ 709, 91191 Gif sur Yvette, France\\
$^9$University College London, Mullard Space Science Laboratory, Holmbury St.\ Mary, Dorking, Surrey RH5 6NT, UK\\
$^{10}$Los Alamos National Laboratory, Los Alamos, New Mexico 87545, USA\\
$^{11}$NASA Goddard Space Flight Center, Greenbelt, Maryland 20771, USA\\
$^{12}$ Universities Space Research Association, 10211 Wincopin Circle, Suite 500, Columbia, MD 21044\\
$^{13}$Universit\`a degli Studi di Padova, Dipartimento di Fisica, via F.~Marzolo 8, 35131 Padova, Italy \\
$^{14}$ INAF/Osservatorio Astronomico di Brera, Via Bianchi  46, 23807 Merate (Lc), Italy\\
$^{15}$ JILA, University of Colorado, Boulder, CO 80309-0440, USA
}
\date{Accepted 2010 May 11.  Received 2010 April 9; in original form 2009 October 21}

\pagerange{\pageref{firstpage}--\pageref{lastpage}} \pubyear{2009}

\begin{document}

\label{firstpage}
\maketitle
\begin{abstract}

We report on the detailed study of the 2008 October outburst from
the anomalous X-ray pulsar (AXP) \src\ discovered through the
\swift/Burst Alert Telescope (BAT) detection of SGR-like short X-ray bursts 
on 2008 October 3. The \swift/X-ray Telescope (XRT) started observing the source 
after less
than 100\,s since the BAT trigger, when the flux
($\sim$$6\times10^{-11}$ erg cm$^{-2}$ s$^{-1}$ in the 2--10 keV range) was
$>$50 times higher than its quiescent level. \swift\ monitored the
outbursting activity of \src\ on a daily basis for
approximately three weeks. This strategy allowed us to find a
phase-coherent solution for the source pulsations after the burst,
which, besides $P$ and $\dot{P}$, requires a positive  $\ddot{P}$ term (spin-down increase). The time evolution of the pulse shape is complex
and variable, with the pulsed fraction increasing from 20\% to
50\% within the \swift\ observational window. The XRT spectra can
be fitted well by means of a single component, either a power-law
(PL) or a blackbody (BB). During the very initial phases of the
outburst the spectrum is hard, with a PL photon index $\Gamma\sim$
2 (or $kT\sim 1.4$\,keV) which steepens to $\Gamma\sim$4  (or
$kT\sim 0.8$\,keV) within one day from the BAT trigger, though the
two components are likely present simultaneously during the first
day spectra. An \int\ observation carried out five days after the
trigger provided an upper limit of $\sim$$2\times10^{-11}$ \flux\ to
the emission of \src\ in the 18--60\,keV band.

\end{abstract}

\section{Introduction}

Magnetars are a small class of isolated X-ray pulsars, believed to have
the highest magnetic fields known to date ($\sim$$10^{14}$--$10^{15}$\,G;
\citealt{duncan92,thompson93}). This class comprises the Anomalous X-ray Pulsars
(AXPs) and the Soft Gamma-ray Repeaters (SGRs), observationally very similar in
many respects (see \citealt{mereghetti08} for a recent
review): a spin period in the 2--12\,s range, large period derivatives
($10^{-13}$--$10^{-10}$ \ss), unpredictable bursting activity on
different timescales (from ms to hundreds of seconds) and luminosities
($10^{38}$--$10^{46}$ \lum). Until not long ago AXPs were thought as
persistent and stable X-ray sources. Only in 2003 the first transient
AXP was discovered by \rxte, namely \xte, which displayed a factor of $>$100
flux enhancement with respect to the pre-outburst 
luminosity level as seen by the \emph{ROSAT} and \emph{Einstein} missions ($\sim$$10^{33}$ \lum;
\citealt{ibrahim04,israel04, gotthelf04,gotthelf05,bernardini09}).
%Unfortunately, the initial phases of the outburst were missed and no
%information on the outburst onset are available.

Even more surprising was the discovery of a highly variable pulsed radio
emission which followed the \xte\ outburst \citep{camilo06,helfand07}, never
observed before in any other magnetar \citep{burgay06}. Moreover, the transient
nature of this AXP provided the first hint that a relatively large number of members
of this class has not been discovered yet, and suggested that others would manifest
themselves in the future through a phenomenology (outburst) similar to that
displayed by \xte.

Indeed, after this first discovery, two more transient AXPs have been detected
showing similar outbursts, \wes, \citep{muno07,icd07} and \src.
The latter AXP was first suggested as a candidate magnetar in the supernova
remnant G327.24--0.13 through X-ray observations \citep{gelfand07}, and
subsequently recognized as a radio transient magnetar through the
discovery of radio pulsations \citep{camilo07a,camilo08} at a period of 
$\sim$2.1\,s, and of an X-ray outburst \citep{halpern08}. 
In particular, \swift/X-ray Telescope (XRT) observations taken
in Summer 2007 revealed \src\ at an X-ray flux level of
$\sim$$5\times10^{-12}$ erg cm$^{-2}$ s$^{-1}$ (more than one order of magnitude brighter
than in quiescence),  with characteristics similar to those of \xte\ in
outburst. No X-ray bursts were observed during the 2007 outburst of
\src, possibly due to a sparse X-ray coverage. More generally, the first phases
of the 2003 and  2007 X-ray outbursts from \xte\ and \src, respectively, were
missed.

Relatively deep \xmm\ pointings were obtained in August 2006 and 2007, the first during
quiescence, and the second during the outburst decay. Pulsed fractions of about
$\sim$15\% and $\sim$7\% were inferred for the 2006 and 2007 observations,
respectively \citep{halpern08}. The 2006 quiescent spectrum
($\sim$$4\times10^{-13}$ erg cm$^{-2}$ s$^{-1}$ in the 1--10 keV range)
can be fitted well
with the usual thermal (blackbody) plus non-thermal (power-law) model ($kT=0.40$ keV and $\Gamma=3.2$; \citealt{halpern08}), {\rc with a 1D resonant cyclotron scattering model (with a surface temperature $kT$=0.33 keV, an electron velocity $\beta$=0.32 and an optical depth $\tau$=1.0; \citealt{rea08}), or with 3D Monte Carlo model  (with a surface temperature of $kT$=0.33 keV, an electron bulk velocity of $\beta_{bulk}$=0.15, and a twist angle of $\Delta\phi$=1.14; \citealt{zane09})}. On the other hand, the 2007  spectrum ($\sim$$3\times10^{-12}$ erg cm$^{-2}$ s$^{-1}$ in the 1--10 keV range)  is characterized by a harder emission with $kT=0.52$ keV and $\Gamma=1.8$  (by using the BB plus PL model). 

Deep infrared observations taken from ESO--VLT during the 2007 outburst have
revealed four objects consistent with the radio AXP position, although none of
them showed variability \citep{mignani09}. Recently,
during the giant outburst detected from \src\ on 2009 January 22, a relatively
bright transient IR source ($K_{\rm{s}}\sim18.5$ mag) was discovered within the 
radio positional uncertainty region of the source \citep{Israel09a}.

Within the magnetar scenario, AXP outbursts are thought to be caused
by large scale rearrangements of the external magnetic field,
either accompanied or triggered by fracturing of the neutron-star crust.
%which might be caused by (or origin of) the observed short X-ray bursts.
These events may result in renewed magnetospheric activity
through the interaction between thermal surface photons and
charges flowing along the (closed) field lines. In addition, hot spots on the 
neutron-star surface may appear where the currents impact on the star. Repeated resonant
cyclotron scatterings onto the magnetospheric particles result in a
modification of the seed thermal spectrum with the appearance of a non-thermal,
high-energy tail \citep{tlk02}.

During the first month since the onset of an outburst, 
the source experience a large flux variations (approximately by one
or two orders of magnitudes). In this respect, monitoring campaigns,
obtained just after the outburst onset, are giving an important opportunity
to track the evolution of the main physical parameters and, therefore,
to check the goodness of the (empirical and/or more physical)
components used to model the spectra.
Here we report on the discovery of bursting activity and a new
outburst from \src, as observed by \swift\ (\S\ref{swift}), which was
promptly triggered by the onset of the bursting activity of this
transient AXP on 2008 October 3. The \swift\ monitoring of the source in the
100\,s--22\,days time interval allowed us to carry out a detailed
timing and spectral study, the first ever for this source, of the initial phases
of its outburst on a daily basis. We also report on the results of an
\int\ Target of Opportunity observation performed during the X-ray outburst (\S\ref{integral}), and discuss our findings in comparison with other magnetar outbursts (\S\ref{disc}).

\section{\emph{Swift} observations and data analysis}
\label{swift}

The \swift\ satellite \citep{gehrels04short} is an efficient
observatory for the discovery and the multiwavelength monitoring of
AXP bursting activity.  Whenever the wide-field coded-aperture mask
Burst Alert Telescope (BAT; \citealt{barthelmy05}) triggers on a burst
from an interesting hard X-ray transient, the system on-board can
rapidly re-point the satellite in order to bring the source in the
narrower field of view of the other two telescopes on-board, the X-Ray
Telescope (XRT; \citealt{burrows05short}) and the Ultraviolet/Optical
Telescope (UVOT; \citealt{roming05short}). In this section we present
the results obtained from our analysis of the \swift\ BAT and XRT
observations of AXP \src\ performed during its bursting activity on
2008 October, being the source too absorbed to be detected with the UVOT.

\subsection{Burst Alert Telescope data}
\label{bat}
\begin{figure*}
 \begin{minipage}[t]{\hsize}
\resizebox{\hsize}{!}{\includegraphics[angle=-90]{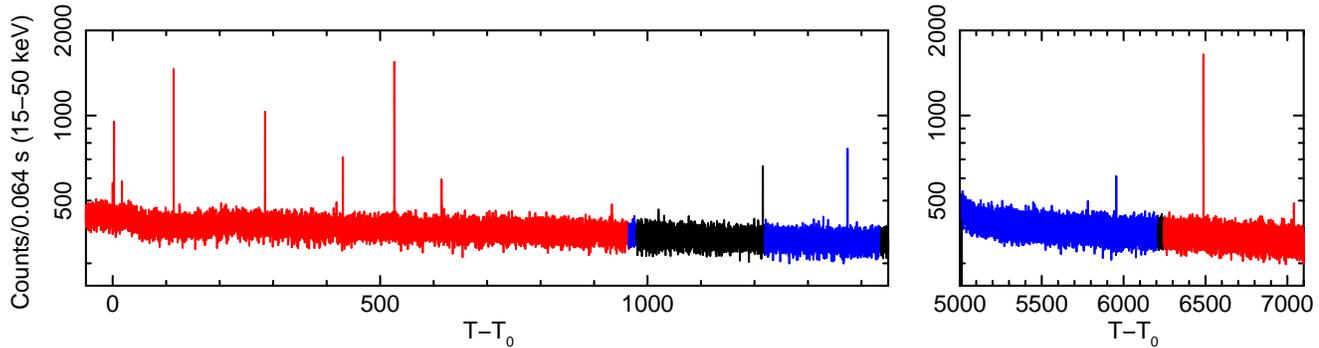}}
\caption{\label{BAT_timeline}The light curve of \src\ generated on-board by the
\swift\ Burst Alert Telescope (rate data). Black correspond to temporal
ranges for which only rate data were available, red and blue to the temporal
ranges for which also event and survey data, respectively, were distributed.$T_{\circ}$ corresponds to the first BAT trigger (trigger 330353; 2008-10-03 09:28:08 \textsc{ut}). The units of the horizontal axis variable are seconds.}
\label{BAT_LC}
 \end{minipage}
\end{figure*}

On 2008 October 3 at 09:28:08 \textsc{ut} BAT triggered on and localized a
short burst from a position consistent with that of the AXP \src\ (trigger
330353; \citealt{krimm08_gcn8311}) which
was followed by several other bursts extending to at least 09:38:24 \textsc{ut}.
About 2 hours later, at 11:16:13 \textsc{ut}, BAT detected one last, bright burst
which reached a peak count rate of $\sim$20\,000 counts s$^{-1}$ in the 15--350
keV band (trigger 330367;  \citealt{krimm08_gcn8312}).
%On 3 October 2008 at 09:28:08 UT BAT triggered and localized a short burst from AXP 1E1547.0-5408 \citep{GCN8311} which was followed by several other bursts extending to at least 09:38:24 UT. About 2 hours later, at 11:16:13 UT, BAT detected a last, bright burst which reached a peak count rate of $\sim$20000 count/s in the 15-350 keV band \citep{GCN8312}.
The BAT time coverage of the source was discontinuous along the bursting period,
totalling an effective exposure time of 4 ks (see Figure \ref{BAT_timeline}).

%The BAT time coverage of the source was discontinuous along the bursting period,
%totaling a net exposure time on-target of 4 ks (see Figure \ref{BAT_timeline}). Both data
%containing information for individual photons (`event data') and data composed of
%80-channel histograms with a typical integration time of $\sim$5 minutes (`survey
%data') were distributed. Since no bursts were detected in survey data, we present
%in this section the results of our analysis of the bursts detected in the event
%data (net exposure-time of 2.1 ks).
Datasets containing information for individual photons (event data) and data
composed of 80-channel histograms with a typical integration time of $\sim$5
minutes (survey data) were distributed.
%In the survey data, covering $\sim$1.9 ks of the total exposure time (see the
%blue fraction of the light curve shown in Figure \ref{BAT_timeline}), we did
%not detect any persistent emission from the source (we inferred a 3$\sigma$
%upper limit of XXX ergs\,cm$^{-2}$\, s$^{-1}$ in the  XX-XXkeV range).
In the following, we present the results of our analysis for the bursts detected
in the event data.
\begin{figure*}
\begin{minipage}[t]{.8\hsize}
\resizebox{\hsize}{!}{\includegraphics[angle=-90]{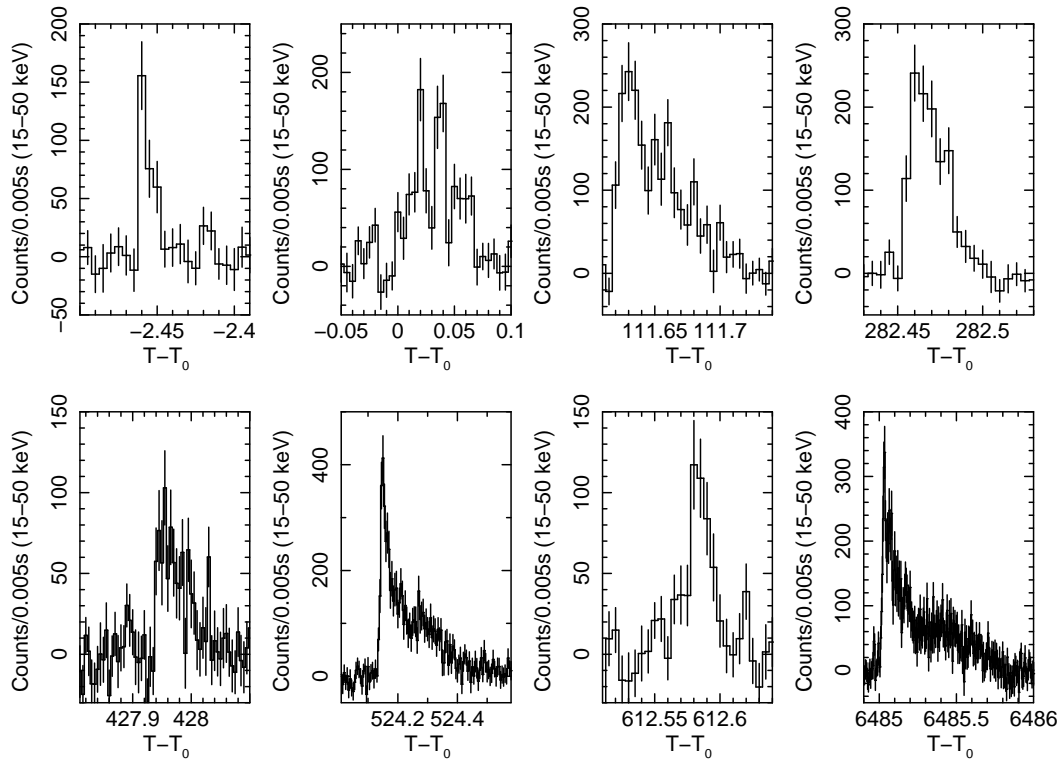}}
%\resizebox{\hsize}{!}{\includegraphics[angle=-90]{phases_1e1547.ps}}
\caption{\swift/BAT background subtracted light curves of the \src\ bursts. Time bins are of 5 ms. The units of the horizontal axis variable are seconds. {\rc Negative times marks  the not triggered burst few seconds before T$_0$.}\label{bat_lcurve} }
\label{BATbursts}
\end{minipage}
\end{figure*}

\begin{table*}
\centering
\begin{minipage}{15.cm}
\caption[]{Spectral fit results (in the 15--100 keV range) of the bright bursts
detected by \swift/BAT for the BB model. We assume a distance to the source of 4
kpc. $T_{0}$ corresponds to the first BAT trigger (2008-10-03 09:28:08) and errors are reported at 1$\sigma$ confidence level.}
\label{Table_BAT_spec_fit}
\begin{tabular}{@{}lcccccccc}
\hline

Burst time (T--T$_{0}$)  & $\tau$                    & Net counts  &  Duration &
$kT$                       & R                           &  L
      &  Fluence   &  $\chi^{2}_{r}$(d.o.f.) \\
                        & (s)                         &             &    (s)
& (keV)                    & (km)                        &  erg s$^{-1}$
       &  erg cm$^{-2}$  &   \\
\hline \\

-2.45           & 0.007$_{-0.002}^{+0.002}$ & 358         &  0.016    &
10.1$_{-1.0}^{+0.9}$     &     1.5$_{-0.3}^{+0.3}$     &     6.0 $\times$
10$^{38}$      & 4.4 $\times$ 10$^{-9}$   & 0.93(38) \\
0               & 0.018$_{-0.004}^{+0.005}$ & 1492        &  0.070    &
10.9$_{-0.8}^{+0.8}$     &     1.3$_{-0.2}^{+0.2}$     &     6.4 $\times$
10$^{38}$      & 2.1 $\times$ 10$^{-8}$   & 1.04(38) \\
111.6           & 0.036$_{-0.003}^{+0.003}$ & 2964        &  0.098    &
12.4$_{-0.5}^{+0.6}$     &     1.2$_{-0.1}^{+0.1}$     &     9.4 $\times$
10$^{38}$      & 4.3 $\times$ 10$^{-8}$   & 1.18(38) \\
282.45          & 0.015$_{-0.001}^{+0.001}$ & 1451        &  0.035    &
11.3$_{-0.7}^{+0.8}$     &     1.6$_{-0.2}^{+0.2}$     &     1.1 $\times$
10$^{39}$      & 1.9 $\times$ 10$^{-8}$   & 1.44(38) \\
427.9           & 0.038$_{-0.007}^{+0.009}$ & 988         &  0.100    &
 9.3$_{-1.0}^{+1.1}$     &     1.2$_{-0.2}^{+0.3}$     &     2.8 $\times$
10$^{38}$      & 1.3 $\times$ 10$^{-8}$    &  0.88(38) \\

524.2           & 0.038$_{-0.005}^{+0.007}$ & 8081        &  0.254
&  11.8$_{-0.3}^{+0.3}$     &     1.4$_{-0.1}^{+0.1}$     &     1.0 $\times$
10$^{39}$      & 1.2 $\times$ 10$^{-7}$   &  0.80(38) \\
                        & 0.133$_{-0.008}^{+0.009}$ &             &            &
                         &                             &
        &                        &           \\
524.2 rise      &                           & 310         &  0.010
&  11.1$_{-1.0}^{+1.1}$     &     1.5$_{-0.3}^{+0.3}$     &     8.6 $\times$
10$^{38}$      & 4.0 $\times$ 10$^{-9}$    &  1.34(38) \\
524.2 peak      &                           & 742         &  0.007
&  8.3$_{-0.4}^{+0.4}$      &     4.6$_{-0.5}^{+0.5}$     &    2.6 $\times$
10$^{39}$      & 8.0 $\times$ 10$^{-9}$     &  1.01(38) \\
524.2 decay     &                           & 7029        &  0.237    &
11.9$_{-0.3}^{+0.3}$     &     1.3$_{-0.1}^{+0.1}$     &     9.4 $\times$
10$^{38}$      & 1.0 $\times$ 10$^{-7}$    &  0.94(38) \\

612.55          & 0.012$_{-0.002}^{+0.003}$ & 757         &  0.100    &
11.1$_{-1.7}^{+1.9}$     &     0.7$_{-0.2}^{+0.3}$     &     2.0 $\times$
10$^{38}$      & 9.5 $\times$ 10$^{-9}$    &  0.92(38) \\

6485            & 0.098$_{-0.004}^{+0.004}$ & 14317       &  0.730    &
10.8$_{-0.3}^{+0.3}$     &     1.3$_{-0.1}^{+0.1}$     &     5.8 $\times$
10$^{38}$      & 2.0 $\times$ 10$^{-7}$   &  1.17(38) \\
                        & 0.283$_{-0.018}^{+0.019}$ &             &            &
                         &                             &
        &                        &           \\
6485 rise           &                           & 676         &  0.020    &
9.6$_{-0.9}^{+1.0}$      &     2.0$_{-0.3}^{+0.4}$     &     8.6 $\times$
10$^{38}$      & 8.0 $\times$ 10$^{-9}$    &  0.82(38) \\
6485 peak           &                           & 806         &  0.010    &
7.4$_{-0.3}^{+0.3}$      &     5.0$_{-0.5}^{+0.5}$     &     2.0 $\times$
10$^{39}$      & 8.2 $\times$ 10$^{-9}$    &  1.03(38) \\
6485 decay          &                           & 12835       &  0.700    &
10.9$_{-0.3}^{+0.3}$     &     1.2$_{-0.1}^{+0.1}$     &     5.6 $\times$
10$^{38}$      & 1.8 $\times$ 10$^{-7}$   &  1.08(38) \\

\hline
\end{tabular}
\end{minipage}
\end{table*}
%
%The data reduction was performed using version 2.9 of the standard \swift/BAT
%software distributed within \textsc{ftools} under the \textsc{heasoft} package
%(version 6.5). A weight map, taking into account the contribution of the source
%to the overall detector counts (mask-weighting technique),\footnote{See\\
%http://heasarc.nasa.gov/docs/swift/analysis/threads/batfluxunitsthread.html.}
%was applied in order to obtain the background-subtracted counts of the source.
%In Figure \ref{BAT_timeline} the light curve in the 15--50 keV band showing the bursting
%activity is plotted. Whenever the burst had sufficient statistics, time-resolved
%spectra were generated corresponding to the rise, peak, and decay phases of
%the burst. Otherwise a spectrum of the entire burst was extracted.\\

The data reduction was performed using version 2.9 of the standard
\swift/BAT software distributed within \textsc{ftools} under the
\textsc{heasoft} package (version 6.5.1). A weight map, taking into account the
contribution from the source to the counts in each BAT detector (mask-weighting
technique),\footnote{See\\
http://heasarc.nasa.gov/docs/swift/analysis/threads/batfluxunitsthread.html.}
was applied in order to obtain the background-subtracted counts of the source.

In Figure \ref{bat_lcurve} the 15--50 keV light curves of bursts with
more than 350 counts are plotted. For each burst we extracted a spectrum.
All spectra are fitted well by a single blackbody model (BB) with a
temperature of about 11 keV and an emitting radius of about 1 km (see Table
\ref{Table_BAT_spec_fit}). The limited statistics of BAT spectra did not allow 
us to check whether an additional BB component is present similar to the case of 
SGR\,1900+14 \citep{israel08short}.
We also tried to fit the {\rc 15-100 keV} spectra with a power law and an optically thin thermal
bremsstrahlung obtaining worse {\rc reduced} $\chi^2$ values (in the 1.4$\div$1.6 
and 1.7$\div$2.9 range for the faintest and brightest bursts, respectively, {\rc and for 38 degree of freedom}).
The characteristic time-scales $\tau$ of each burst decay was also inferred by fitting 
the corresponding light curves with one or two (for the longest events) exponential laws  and are reported in Table \ref{Table_BAT_spec_fit}.\\
%Light curves of the bursts in the 15-30 keV and 30-80 keV energy range were also
%produced in order to perform a hardness ratio.
\indent Whenever the bursts had sufficient statistics,  time-resolved spectra were
generated corresponding to the rise, peak, and decay phases of the
bursts. The results of the fitting procedure is reported in
Table\,\ref{Table_BAT_spec_fit}. No significant variations were found
in the spectral parameters, except for the normalization
factor due to the luminosity evolution (decay) during the burst. On the other hand,  
time-resolved spectra show a moderate spectral evolution with the peak BB temperature being 
systematically softer than that during decay (similar results where obtained for \sgra\ and \sgrb: \citealt{diego04,diego06,israel07}).\\
%\indent The presence of a significant persistent emission over all the
%non-bursting time
%period was also searched for in the event data. An image was generated from all the events
%of each file covering time intervals where no bursts were present and the standard source
%detection was run (\textsc{batcelldetect}). Two time intervals were investigated, the first
%one ranging from $t=-41.5$ s to $t=965.5$ s (net exposure time 1001\,s) and the second one
%from $t=6248.5$ s to $t=7300$ s (net exposure time 1050\,s). No significant emission in the
%direction of AXP \src\ was found. The 3$\sigma$ upper limits on the flux in the 14--50 keV
%band for the two intervals described above are $6\times10^{-10}$ and $2\times10^{-10}$ \flux,
%respectively.
\indent The presence of persistent emission during the non-bursting time intervals 
was also searched for in the event data. An image was generated from all the
events of each file covering time intervals where no bursts were present and the
standard source detection was run (\textsc{batcelldetect}). Two time intervals
were investigated, the first one ranging from $t=-41.5$ s to $t=965.5$ s (net
exposure time 1\,001 s) and the second one from $t=6\,248.5$ s to
$t=7\,300$ s (net exposure time 1\,050 s). No significant emission in the direction
of AXP \src\ were found. The 3$\sigma$ upper limits on the flux in the
14--50 keV band for the two intervals described above are $6\times10^{-10}$ and
$2\times10^{-10}$ \flux, respectively.

\subsection{X-Ray Telescope data}
\label{table:xrt}
%%%%%%%%%%%%%%%%%%%%%%%%%%%%%%%%%%%%
\begin{table*}
\centering
\begin{minipage}{13.3cm}
\caption{\swift/XRT observation log of \src.}
\label{log}
\begin{tabular}{@{}ccccrr}
\hline
Sequence & Mode & Start time (\textsc{ut}) & End time (\textsc{ut}) & Exposure$^{\rm
a}$ & Time since trigger\\
 & & yyyy-mm-dd hh:mm:ss & yyyy-mm-dd hh:mm:ss & (ks) & (d)\\
\hline
\phantom{$^{\rm b}$}00330353000$^{\rm b}$ & PC & 2008-10-03 09:29:47 &
2008-10-03 12:52:41 & 3.9 & 99\,s \\
00330353001 & WT & 2008-10-03 12:54:35 & 2008-10-04 01:39:47 & 14.2 & 0.14
\\
00330353002 & WT & 2008-10-04 17:27:08 & 2008-10-04 20:53:52 &
\phantom{1}4.8 & 1.33 \\
00330353004 & WT & 2008-10-05 09:42:43 & 2008-10-06 21:10:00 &
\phantom{1}10.5 & 2.01  \\
00330353005 & WT & 2008-10-07 06:39:12 & 2008-10-07 18:15:00 &
\phantom{1}7.6 & 3.88 \\
00330353006 & WT & 2008-10-08 00:20:12 & 2008-10-08 05:13:00 &
\phantom{1}4.5 & 4.62 \\
00330353007 & WT & 2008-10-09 05:20:33 & 2008-10-09 13:30:31 &
\phantom{1}1.0 & 5.83 \\
00330353008 & WT & 2008-10-10 07:05:50 & 2008-10-10 15:16:00 &
\phantom{1}3.9 & 6.90 \\
00330353010 & WT & 2008-10-12 07:11:17 & 2008-10-12 18:31:00 &
\phantom{1}3.7 &  8.90 \\
00330353011 & WT & 2008-10-13 04:05:15 & 2008-10-13 15:23:00 &
\phantom{1}3.4 &  9.77 \\
00330353012 & WT & 2008-10-16 01:05:20 & 2008-10-16 06:09:00 &
\phantom{1}4.0 &  12.65 \\
00330353013 & WT & 2008-10-17 23:38:20 & 2008-10-18 14:13:00 &
\phantom{1}3.7 & 14.59 \\
00330353014 & WT & 2008-10-20 14:47:50 & 2008-10-21 00:00:00 &
\phantom{1}3.9 & 17.22 \\
00330353015 & WT & 2008-10-22 08:03:14 & 2008-10-22 13:08:00 &
\phantom{1}3.9 & 18.94 \\
00330353016 & WT & 2008-10-24 01:56:29 & 2008-10-24 08:33:59 &
\phantom{1}3.6 & 20.68 \\
\hline
\end{tabular}
\begin{list}{}{}
\item[$^{\rm a}$] The exposure time is usually spread over several snapshots
(single continuous pointings at the target) during each observation.
\item[$^{\rm b}$] During this observation XRT repeatedly switched between PC and
WT modes; since the exposure in WT is short (about 39\,s), we did not include it in our analysis.
\end{list}
\end{minipage}
\end{table*}
%%%%%%%%%%%%%%%%%%%%%%%%%%%%%%%%%%%%
\swift\ executed a prompt slew on \src\ and the first XRT
observation started only 99 s after the BAT trigger. Table \ref{log} reports the log
of the observations that were used for this work. The XRT uses a CCD detector 
sensitive to photons with energies between 0.2 and 10 keV. We considered Windowed 
Timing (WT) and Photon counting (PC) mode data. In PC
mode the entire CCD is read every 2.507\,s, while in WT mode only the central 200 columns
are read and only one-dimensional imaging is preserved, achieving a time resolution of
1.766 ms (see \citealt{hill04short} for a detailed description of XRT modes).\\
\indent The data were processed with standard procedures using the \textsc{ftools} task
\textsc{xrtpipeline} (version 0.12.0) and events with grades 0--12 and 0--2 were selected
for the PC and WT data, respectively (see \citealt{burrows05short}). For the timing and
spectral analysis, we extracted the PC source events {\rc in the 0.5-10 keV range} within a circle with a radius of 30
pixels ($\sim$$71\arcsec$). The WT data {\rc (in the 0.5-10 keV range)} were extracted in a rectangular region 40 pixels
long (and 20 pixels wide) along the image strip. To estimate the background, we extracted
PC events within an annular region (radii of 50 and 75 pixels) centered on the source 
and WT events within a rectangular box ($40\times20$ pixels) far from \src.\\
\begin{figure}
% \begin{minipage}[t]{\hsize}
%\resizebox{\hsize}{!}{
\includegraphics[width=7.9cm,angle=-90]{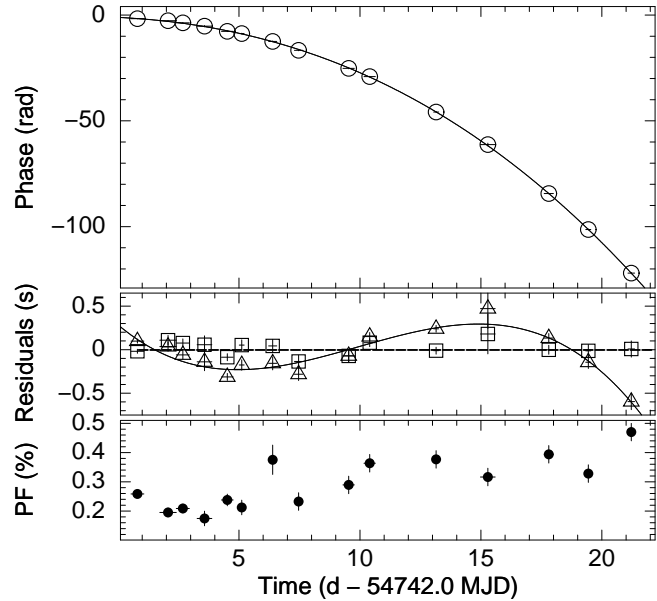}
%}
%\resizebox{\hsize}{!}{\includegraphics[angle=-90]{phases_1e1547.ps}}
\caption{0.5--10\,keV \swift/XRT pulse phase evolution with time,
together with the time residuals (central panel) with
respect to the two phase-coherent timing solutions discussed in the text: $P$--$\dot{P}$
(triangles) and ${P}$--$\dot{P}$--$\ddot{P}$ (squares). The solid lines in the upper panel
represents the latter timing solution. The line in the central panel is the cubic fit to the ${P}$--$\dot{P}$ solution. The pulsed fraction (defined as the semi-amplitude of modulation divided by the mean source count rate) evolution is reported in the lower
panel (where 0.2 means 20\% pulsed fraction).}
% \end{minipage}
\label{figure:phase}
\end{figure}
%%%%%%%%%%%%%%%%%%%%%%%%%%%%%%%%%%%%%
\begin{figure}
\resizebox{\hsize}{!}{\includegraphics[angle=-90]{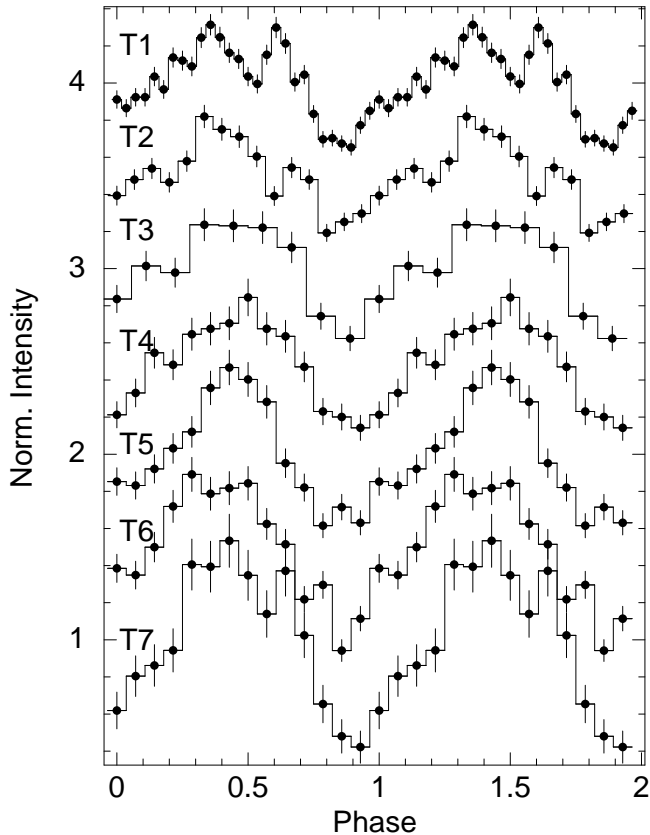}}
\caption{Time evolution of the 0.5--10 keV light curves folded by means of the
${P}$--$\dot{P}$--$\ddot{P}$ phase-coherent timing solution. Each
light curve is covering approximately a three days time interval from the
BAT trigger, where T1 to T7 correspond the the first and last time interval. Light curves are arbitrarily shifted along the vertical axis.}
\label{figure:efold}
\end{figure}
%%%%%%%%%%%%%%%%%%%%%%%%%%%%%%%%%%%%%
\subsubsection{Timing}
\label{sec:timing}
The {\rc 0.5-10 keV events} were used to study the timing properties of the pulsar after
having corrected the photon arrival times to the barycentre of the Solar system
with the \textsc{barycorr} task {\rc (RA=15$^h$ 50$^m$ 54$\fs$12, Dec=-54$^{\rm o}$ 18$^{\prime}$ 24$\farcs$19 and J2000 was assumed for the source position; Israel et al. 2010)}. We started by inferring an accurate period
measurement by folding the data from the first XRT pointing (in WT mode) at the
period reported by \citet{camilo07a} and studying the phase evolution within the
observation by means of a phase-fitting technique (details on this technique are
given in \citealt{dallosso03}). {\rc Due to the high variability of the pulse shape we 
decided not to use a pulse template in the cross-correlation (which might artificially 
affect the phase shift), but to fit each individual folded light curve with the fundamental plus two higher harmonics. In the following we also implicitly assumed that the pulsation period  (and its derivatives) is a reliable estimate of the spin period (and its derivatives), an assumption which is usually considered correct for isolated neutron stars.} The resulting best-fit period {\rc ($\chi^2$=1.1 for 2 degree of freedom, hereafter d.o.f.)} is $P=2.071\,336(3)$\,s (1$\sigma$ c.l.; epoch 54742.0 MJD; see \citealt{esposito08_atel}). The above
period accuracy of 3 $\mu$s is enough to phase-connect coherently the later
\swift\ pointings carried out on a daily baseline (see Table\,\ref{log}). Therefore,
the procedure was repeated by adding, each time, a further \swift\ pointing. The
relative phases were such that the signal phase evolution could be followed unambiguously
for the whole \swift\ visibility window (see Figure \ref{figure:phase}). A quadratic
term was clearly present in the phase time evolution since the very beginning. The
resulting phase-coherent solution had a best-fit period $P=2.071\,336\,8(3)$\,s and
period derivative $\dot{P} = 4.52(4)\times 10^{-11}$\,s s$^{-1}$ (MJD 54742.0
was used as reference epoch; 1$\sigma$ c.l.). These values are in agreement (within $\sim$3$\sigma$) with those, $P=2.071\,335(1)$ s and $\dot{P}=5.7(5)\times 10^{-11}$\,s s$^{-1}$, reported by \citet{israel08atel} and based on a reduced \swift\ dataset. The fit of the phases with a linear and a quadratic term remained statistically acceptable (reduced
$\chi^2_r \sim 0.8\div1.2$ {\rc for $8\div9$ d.o.f.}) for the first two weeks since the BAT trigger. However, during the
latest week the pulse phases increasingly deviate from the extrapolation of the above
$P$--$\dot{P}$ solution (see Figure\,\ref{figure:phase} central panel, empty triangles),
resulting in an unacceptable fit  ($\chi^2_r \sim 18$ for 12 d.o.f.).  Therefore, we added a higher order component 
in order to take into account for the possible presence of a temporary or
secular $\ddot{P}$ term. The resulting new phase-coherent solution is  $P=2.071\,340\,1(7)$ s, $\dot{P} = 3.2(2) \times 10^{-11}$ s
s$^{-1}$, and $\ddot{P} = 2.0(4) \times 10^{-17}$ s s$^{-2}$ (MJD 54\,742.0 was
used as reference epoch; 1$\sigma$ c.l.; $\chi^2_r = 2$ for 11 d.o.f.), or
$\nu = 0.482\,779\,2(2)$ Hz, $\dot{\nu} = -7.4(5) \times 10^{-12}$ Hz s$^{-1}$,
and $\ddot{\nu} =-4.8(9) \times 10^{-18}$ Hz s$^{-2}$. The time residuals with
respect to the new timing solution are reported in Figure\,\ref{figure:phase}
(central panel; empty squares). The significance level for the inclusion of the
cubic component is 5.1$\sigma$.
%\\\textbf{Valutata con F-test? Credo che non sia lecito in questo caso: visto
%che il miglioramento del $\chi^2$ \`e evidente, forse non \`e necessario
%quantificare. Toglierei l'ultima frase. --- P.\\
%Giallo: questo e' il tipico caso in qui l'Ftest va bene.... questo perche' la
%null hypothesis per Pdotdot non e' 0. Infatti Pdotdot puo' assumere valori da
%+infinito a -infinito}\\
Moreover, the new timing solution implies a r.m.s. variability of only 0.08
s, {\rc corresponding to a timing noise level of about 3\%, usually observed in isolated 
neutron stars}. {\rc It is worth emphasizing that the second period derivative we found is very unlike to be related to a random change of the pulse profiles which is expected to introduce only a random distribution of the phase residuals, rather then a cubic term. Recent studies on a sample of 366 radio pulsars  showed  that cubic terms on phase 
residuals are possible though smaller by several orders of magnitudes than that detected in \src\ and recorded on time-scales longer (years) than those we are sampling in our  dataset \citep{hobbes10}. More in general the (long-term) timing noise of young radio pulsars 
($t<10^5$yr) can be explained as being caused by the recovery from previous glitch events \citep{hobbes10}.  The latter results make unlike that the period second derivative is due to random noise. Finally, we note that in the above analysis we assumed that the emitting region is fixed in time with respect to the observer, as suggested by studies on other transient magnetars (see \citealt{perna08,albano10}). Correspondingly, we can also reasonably exclude that the cubic term is introduced by a random  motion of the hot-spot.}\\

\indent Based on the above phase coherent solution we also studied the pulse shape and
pulse fraction evolution. These are
changing as shown in Figure\,\ref{figure:phase} (lower panel) and
Figure\,\ref{figure:efold} as a function of time. In particular the pulsed
fraction (semi-amplitude of the sinusoid divided by the mean count rate) has
increased smoothly from 20\% just after the outburst onset, up
to almost 50\% at the end of the \swift\ visibility window, 22 days after the
BAT trigger. The pulse shape is variable too, from a multi-peak profile at the
beginning of the outburst, to a less structured sinusoid during the latest XRT
observations, though the lower statistics of the latter datasets might have
hidden a more complex behavior.
%Within the statistical uncertainties the pulsed fraction is energy independent
%and no signicant pulse shape variations as a function of energy were
%found by dividing the counts in soft and hard energy intervals.
To assess the significance of the observed pulse shape variations as a function
of time, we compared the folded pairs of contiguous lightcurves reported in
Figure\,\ref{figure:efold} by using a two-dimensional Kolmogorov-Smirnov test
\citep{peacock83,fasano87}. The results show that the
probability that they do not come from the same underlying distribution is
$>$9$\sigma$ (T1-T2), $\sim$2.2$\sigma$ (T2-T3), $\sim$3.3$\sigma$ (T3-T4),
$\sim$2.8$\sigma$ (T4-T5), $\sim$3.8$\sigma$ (T5-T6), and $\sim$2.9$\sigma$
(T6-T7).

%\\\textbf{Qui invece la
%significativit\`a delle variazioni si potrebbe quantificare con
%un Kolmogorov-Smirnov bidimensionale --- P.}

\subsubsection{Spectroscopy}
\label{sec:spec}
For the spectral fitting (with \textsc{xspec} version 12.4), the data were
grouped so as to have at least 20 counts per energy bin. The ancillary response
files were generated with \textsc{xrtmkarf}, and they account for different
extraction regions, vignetting and point-spread function corrections. We used
the latest available spectral redistribution matrix (v011) in \textsc{caldb}.

\begin{figure}
\resizebox{\hsize}{!}{\includegraphics[angle=-90]{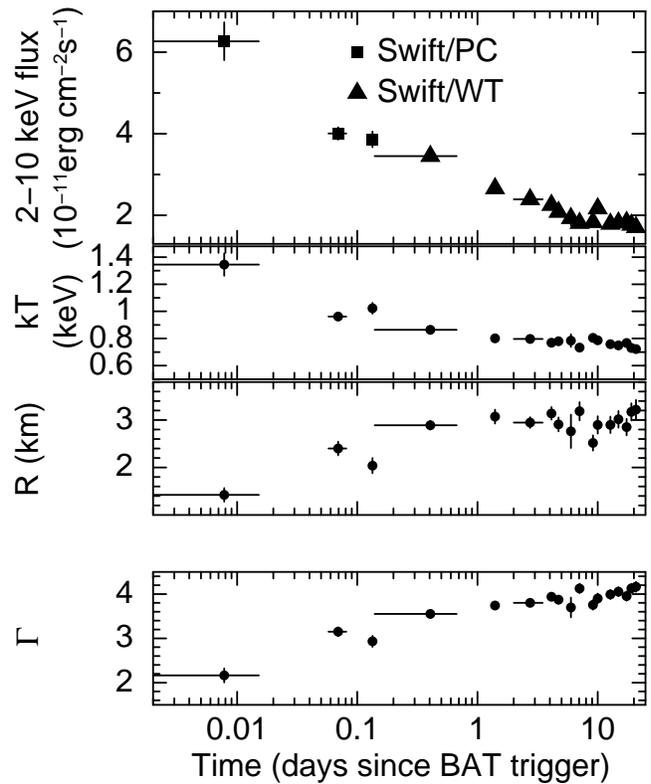}}
%\resizebox{\hsize}{!}{\includegraphics[angle=-90]{unfold.ps}}
\caption{\label{spec_evol} \swift\ XRT 2--10 keV observed fluxes (upper panel)
as a function of the time since the BAT trigger. The two central panels show
the evolution of the characteristic temperature ($kT$) and equivalent radius
($R$) of the blackbody model and inferred from the spectral fitting.
The lowest panel shows the photon index ($\Gamma$) evolution by using a
power-law component.}
\end{figure}

\indent Spectral modelling was performed by fitting together all the datasets
in the 1--10 keV energy range (the high absorption value made the data below
1.5 keV almost useless). We first fit the spectra to simple models: a power law
and a blackbody, both modified for interstellar absorption. All the parameters
were left free to vary, except for the absorption column density, that was
left free but with the request to be the same for all observations. Both models
provide acceptable fits.
The power-law fit ($\chi^2_r = 1.04$ for 1\,292 d.o.f.) yields a column density
$N_{\rm H} = (6\pm1)\times10^{22}$ cm$^{-2}$ and photon indices that increase from
$\Gamma\simeq 2$ to 4 (see Figure \ref{spec_evol}, lower panel). In the data taken
immediately after the first BAT trigger, \src\ was found to be in a record high
state: the observed flux was $\sim$$7\times10^{-11}$ \flux\ in the 2--10 keV band,
about 100 times greater than that observed in 2008 July--September
\citep{esposito08_atel},
and $\sim$10 times greater than the previous highest recorded flux \citep{halpern08}.
The observed flux declined by 70\% in three weeks and the fading trend (see Figure
\ref{spec_evol}, upper panel) can be described by a power-law with decay index
$\alpha=-0.171\pm0.005$ (using the BAT trigger as origin of time).\\
\indent From the blackbody fit ($\chi^2_r = 1.02$ for 1\,292 d.o.f.) we derived a lower
column density of $(2.50\pm0.06)\times10^{22}$ cm$^{-2}$. Again there is a clear indication
of spectral softening during the decay, with the blackbody temperature changing from
 $kT \simeq1.3$ to 0.7 keV (while the radius increased from $\sim$1 to 3 km).

\begin{figure*}
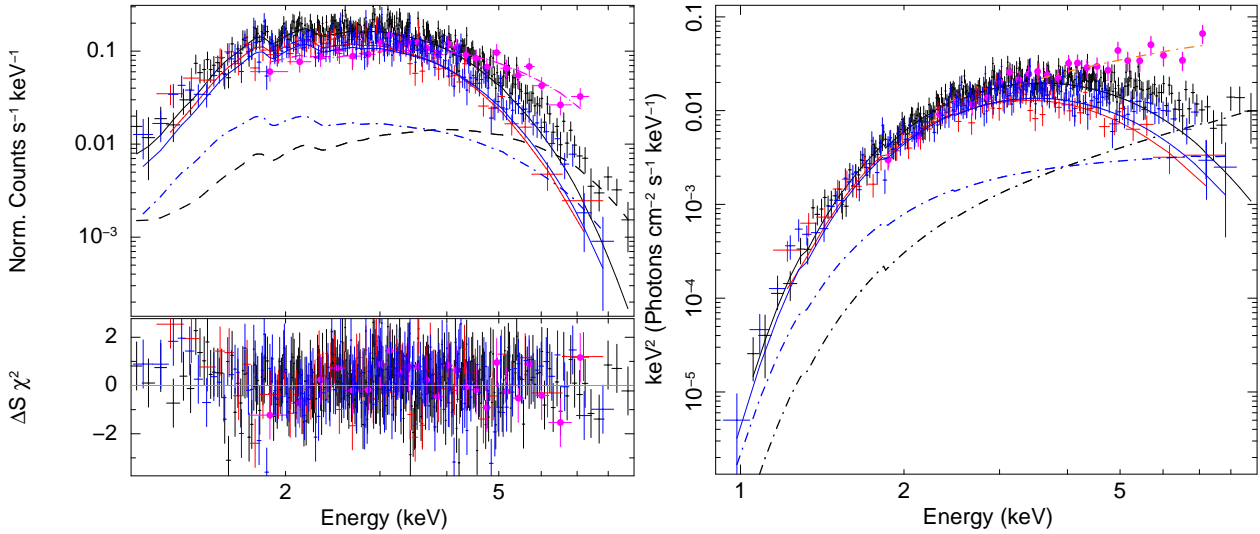

\includegraphics[width=7.cm,angle=-90]{spek_evol.ps}
\includegraphics[width=7.cm,angle=-90]{unfold.ps}
%\resizebox{\hsize}{!}{\includegraphics[angle=-90]{spek_evol.ps}}
%\resizebox{\hsize}{!}{\includegraphics[angle=-90]{unfold.ps}}
\caption{\label{efold} Left panel: \swift\ XRT spectra and residuals of the
first four observations (listed in Table\,\ref{log}) carried out within $\sim$1 
day since the X--ray outburst of \src\ (pink filled circles, black, blue and red 
crosses, from the first to the fourth, respectively). The spectra are fitted by using a blackbody (solid lines) plus a power-law (dot-dash lines) model: the hard-to-soft evolution is evident 
(see text for details). Right panel: same as before but for the density flux in order to
emphasize the spectral evolution above $\sim$4--5\,keV.}
\label{spec}
\end{figure*}

For those observations for which the statistics was good enough (those
obtained during the first day since the BAT trigger), the $\chi^2_r$ was
approximately in the 1.2--1.3 range {\rc (for 549 d.o.f.)}, suggesting a more complex spectrum. In order to check this issue, we re-analyzed the first four {\rc XRT/PC} observations (see Table\,\ref{log}) fitting a BB+PL model. The  results are shown in 
Figure\,\ref{spec} and can be summarized as follows: (a) the first spectrum 
(99\,s after the BAT trigger) can be totally ascribed to a hard power-law 
($\Gamma=1.1\pm0.2$, 90\% c.l.), 
%likely partially
%affected by the presence of five short bursts detected by BAT (see
%Figure\,\ref{BATbursts} and Table\,\ref{Table_BAT_spec_fit}) and not resolved
%within the time resolution of the PC mode, 
(b) during the second and the third
observations (between 0.05 and 0.2 days after the trigger) the BB component
becomes dominant ($kT=0.75\pm0.05$ keV in both cases), though a PL component is 
still needed (but not statistically significant) above 4--6 keV, (c) after one 
day since the BAT trigger (fourth observation) the PL component is not
any more detectable in the BB+PL model, while in the single PL model the photon
index becomes $>$4 (mimicking a Wien tail of a BB component). Given the relatively low number of counts we did not attempt to fit more sophisticated and physical based models.

\section{\emph{INTEGRAL} observation}
\label{integral}

Following the flux increase detected by \swift/XRT we triggered
our \int\ Target of Opportunity observation program.
\int\ \citep{winkler03short} observed the source from
2008 October 8 at 21:35 \textsc{ut} until October 10 at 04:53 \textsc{ut} for
a total exposure time of 95.8 ks. We analyzed the
IBIS \citep{ubertini03short} data, the coded mask imager on board
\int. In particular we selected the data derived from its low energy
(15 keV--1 MeV) detector plane ISGRI \citep{lebrun03}, which is the
most sensitive instrument below 100 keV.

We analyzed the 48 pointings that composed our observation, and
derived the corresponding mosaicked images in two energy bands:
18--60 keV and 18--100 keV. The source is not detected during our
observations, and the 3$\sigma$ upper limits at the source position
are 0.36 and 0.42 counts s$^{-1}$ in the two energy bands respectively.
This corresponds, assuming a power law spectral shape with photon
 index $\Gamma=2.2$, to a flux upper limit of $1.8\times10^{-11}$ \flux\
in the 18--60 keV band, and $2.5\times10^{-11}$ \flux\ in the 18--100~keV
band. Assuming a photon index $\Gamma=1.5$ the upper limits are $2\times10^{-11}$ \flux\
 in the 18--60~keV band and $3\times10^{-11}$ \flux\ in the 
18--100~keV energy band.

In order to search for short bursts, we extracted event lists from
each pointing. The events were then selected in energy (18--60~keV),
and extracted from the pixels that were illuminated by the source direction
for at least 60\% of their surface. From these event lists, we derived light
curves with a binning of 0.05 s. No burst was detected above the background
level. \\

%%%%%%%%%%%%%%%%%%%%%%%%%%%%%%%%%%%%%

%%%%%%%%%%%%%%%%%%%%%%%%%%%%%%%%%%%%%

\section{Discussion and conclusions}
\label{disc}
Thanks to the \swift\ capability of detecting high energy impulsive events with
the BAT, and of rapidly pointing to the BAT position with the narrow field
instrument XRT, we had the
chance of studying the short burst and long-term outbursting properties of one of the
most interesting objects of the  magnetar class, \src, which entered a
renewed phase of activity on 2008 October 3. In particular, given the
relatively large period derivative of the source \citep{camilo07a}, we set up an
observational strategy so that daily pointings were performed during a three
week baseline (during which the source was observable by \swift). We studied
the pulsed emission and inferred a rather accurate phase-connected timing
solution which also includes a second derivative component. The spectral
evolution  was also studied as a function of the decreasing flux level.

The connection between the emission of short X-ray bursts and the increase of
the persistent X-ray emission of AXPs and SGRs, has been reliably assessed in
the last few years, especially thanks to the prompt {\em Swift} X-ray follow-up
observations after the detection of short bursts in \wes, \sgrd, \sgrb\ and
\sgrf\ (\citealt{icd07,eiz08,israel08short,holland08,rea09}). Furthermore,
magnetar outbursts have been found to be characterized by an
intensity-hardness correlation, the harder the more intense (and viceversa), as
also observed in \src. Both the connection with the occurrence of the
bursts, and the hardness intensity correlation point toward a scenario in
which these events are related to the evolution of a twisted magnetic
field, which, once implanted by sudden crustal motions, necessarily decays
\citep{tlk02, belo09}.

Particularly interesting are the burst energetics recorded by BAT. In fact, the
15--150\,keV fluence of the bursts reported in Table\,\ref{Table_BAT_spec_fit}
is in the $\sim 1\times 10^{39}$\,ergs range (inferred for a distance of 4kpc; \citealt{tiengo10}). These estimates are conservative lower limits to the total energy emitted in the bursts 
by the source.  On the other hand, the upper limit on the persistent emission of 
the source, as seen by XRT, is of the order of $\sim 4\times$10$^{38}$\,ergs (by integrating the exponential flux decay law from the beginning of October 2008 
to mid-January 2009; see also first panel of Figure\,\ref{spec_evol}), corresponding 
to a ratio between energy released in the bursts and in the persistent tail $>$2.
The latter value is in between those observed for SGRs, i.e. $>$10, and those of AXPs where the typical ratio is of the order of unity or less.

The X-ray timing properties of \src\ in October 2008 are
such that a strong second derivative component, beside the usual $P$ and $\dot{P}$
ones, is needed in order to account for the evolution of the pulse phases during
the about three weeks covered by \swift\ observations. The latter component is
$\ddot{P} = 2.0(4) \times 10^{-17}$ s
s$^{-2}$ or $\ddot{\nu} =-4.8(9) \times 10^{-18}$ Hz s$^{-2}$, therefore with
the same sign as that of the first derivative. Second period derivatives have
been detected only for a few AXPs/SGRs and, in all cases, occurred in response to
glitches, such as in the case of \rxs\ \citep{dallosso03,israel07,dib08} and \kes\
\citep{dib08}, and/or outbursts,  \xte\ \citep{camilo07b} and \sgrf\
\citep{israel08c,rea09}. In nearly all cases the second derivatives were
relatively small, while the sign of these components was such that it acts in
decelerating the spin down rates; typical $\ddot{\nu}$ values are approximately
in the 0.05$\div$5 $\times$10$^{-21}$\,Hz\,s$^{-2}$ range. Only for one
AXP, namely \rxs, negative second period derivatives in the --0.01$\div$--1.3
$\times$10$^{-20}$\,Hz\,s$^{-2}$ range has been detected just after a glitch
\citep{dib08,israel07}. 

The timing properties of \src\ were already known to be unusual 
during the first 6 months of radio monitoring (June --
November 2007; \citealt{camilo07a}). Even in that case, observations 
were taken after a period of activity and also at that time both the first and
second frequency derivatives were negative:  $\dot{\nu} = -6.5 \times
10^{-13}$ Hz s$^{-1}$ and $\ddot{\nu} = -1.2 \times 10^{-19}$ Hz s$^{-2}$.
Moreover, during the same time span in which the  $\dot{\nu}$ (and torque)  
increased, the X--ray flux decreased, a property which is difficult to 
account for with simple assumptions. According to 
\cite{camilo07a}, a possibility is that an extra torque (either than the 
magnetic spin down) originated from a particle wind that twists the magnetic 
field and, at the same time, transfers some angular momentum. In any case, 
the observed phenomenology calls for a mechanism in which X-ray luminosity 
and spin-down rate are not caused by the same mechanism or, at least, not 
by the same mechanism and also at the same time.  Even more dramatic is 
the case of October 2008 reported in this paper, with \src\ showing a
$\ddot{\nu}$ which is a factor of about 50 larger than in 2007, while the X--ray
flux is a factor of 4--30 higher than during 2007. It is very likely that the
$\ddot{\nu}$ components related with glitch-only events have a different origin
than those related to outburst events. For another transient AXP, \xte, a
$\ddot{\nu}$ component has been detected during an outburst, although after two
years since the onset \citep{camilo07b}. Correspondingly, a straightforward
comparison with the \src\ properties is not feasible (similar arguments also
hold for the radio properties of the two sources). In the case of \sgrf, the
second derivative component was inferred during the first two months since the
outburst onset and has a sign opposite to that of \src. 

A plausible scenario is that the dissimilar behavior seen in 
the various sources is  related to the difference between the outburst
$\dot{P}_{\rm outb}$ and the pulsar secular one $\dot{P}_{\rm sec}$. Regardless of the 
mechanism responsible for the outburst and for the net timing effects at the
outburst onset, it is reasonable to imagine the pulsar as a system which tends
to recover its (almost) initial timing properties, i.e. same pre-outburst and
post-outburst $\dot{P}_{\rm sec}$ {\rc \citep{alpar84a,alpar84b,link92}}. 
Correspondingly, a negative or positive second
period derivatives might be ascribed to whether the $\dot{P}_{\rm outb}$ over
$\dot{P}_{\rm sec}$ ratio is larger or smaller than unity. Unfortunately, for all
the four magnetars for which an outburst was detected in the latest few years (\xte,
\wes, \sgrf\ and \src) the  $\dot{P}_{\rm sec}$ is unknown. In the case of \src\ the
positive second derivative detected in 2007 and 2008 would imply a
$\dot{P}_{\rm outb}$ smaller than $\dot{P}_{\rm sec}$: the latter being the value to
which the source is approaching asymptotically. Within the classical scenario of
isolated neutron stars, $\dot{P}_{\rm sec}$ is the spin-down value to be used when
inferring the pulsar dipolar magnetic field strength.
Correspondingly, the inferred value of $B_d \sim 2.7 \times 10^{14}$ Gauss should 
be considered as a lower limit on the intrinsic dipolar magnetic field strength.
%However, even assuming that $\dot{P}_{sec}$ would be a
%factor of ten lower than $\dot{P}_{outb}$, $B_d$ would be close to $10^{14}$
%Gauss.

Taken at face values, the increase of $\dot P$ during the outburst decay 
appears  difficult to reconcile with the predictions of the twisted
magnetosphere model. In such model, in fact, the spin-down rate is dictated
by the radial variation of the external field. A globally twisted dipolar field
has radial dependence $\sim R^{-2-p}$, where the index $p$ decreases with
increasing twist angle ($0\leq p\leq 1$). If, as it seems natural to assume, the
twist decays as the source returns to quiescence, one expects a decrease
in the spin-down rate, i.e. a negative $\ddot P$. As pointed out by 
\cite{belo09}, however, in realistic situations the twist is likely to be
confined to a rather narrow bundle of current-carrying field lines. In this
case, and if the twist is suddenly implanted but its strength is moderate (twist
angle less than $\sim$1 rad), the twist  may still grow for a while
in spite of the luminosity released by dissipation monotonically decreasing. 
Only after the twist angle has reached its maximum value, it will start
to decrease, together with the torque. This implies that the spin-down rate may
be non-monotonic during the outburst decay, with a first phase characterized by
a positive $\ddot P$ followed by a ``recovery'' during which $\ddot P<0$ and
$\dot P$ goes back to its secular value. The typical duration of the increasing
spin-down stage is difficult to quantify, but it can last $\sim$ months (see
Eq.~55 in  \citealt{belo09}). Indeed, this effect has been already proposed by
\cite{belo09} to explain the anti-correlation between X-ray luminosity and
torque observed in 1E\,1048.1-5937 \citep{gavriil04}. In \src\ 
$\dot P$ has been observed to increase monotonically during summer/fall 2007
\cite[during the decay of an outburst the peak of which has been 
missed;][]{camilo08}. The value of $\dot\nu$ corresponding to the latest radio
observation of \cite{camilo08} is $\sim -7.35 \times 10^{-12}\ {\rm s}^{-2}$,
and that in October 3, at the beginning of the X--ray outburst, $\sim-(7.4\pm 0.4)
\times 10^{-12}$ s$^{-2}$. A continuous decrease in $\dot\nu$ seems
therefore hardly compatible with observations, and, since a constant 
$\dot\nu$ has never been observed in AXPs, it may well be that $\dot\nu$ started
increasing at some epoch between 2007 November 22 (latest radio observations)
and 2008 October 3 (X-ray outburst onset). This behavior would not be in
contradiction with the untwisting magnetosphere model mentioned 
above. Finally, it is worth mentioning the possibility that the unusual 
$\dot P$ behaviour after the October 2008 outburst might be one of the 
reasons for the extremely intense outburst displayed by \src\ in January 2009. 
In other words, the positive $\ddot P$ would be acted in accelerating the onset of 
stresses on the NS surface.  

The pulsed fraction of \src\ clearly increases during October 2008, a
behavior already observed during the initial phases of AXP outbursts, such as
the case of \wes\ where the quiescent pulsed fraction  ($\sim$80\%) dropped to
$\sim$10\% at the outburst onset and  started increasing thereafter
towards the quiescent value (\citealt{icd07}; similar results have been recently
obtained for \sgrf; \citealt{rea09}). However, in the case of \src\ the pulsed
fraction at the outburst onset ($\sim$20\%) is already larger than the quiescent
value (as seen by \xmm; $\sim$7\%) and it reached $\sim$50\% at the end of the
October 2008 \swift\ monitoring window. This finding, together with the very
variable pulse profile as a function of time, suggests that the emission and/or
magnetospheric geometry is complex and variable during the initial phases (on
month timescales) of the outburst, and likely not easily related to the
geometry in quiescence.

Pulsations in the X-ray flux may be a consequence of emission from a limited
area of the star surface, of non-isotropic magnetospheric emission, or a
combination of both effects. Again, if the star surface is heated by the 
returning currents which appear when a magnetospheric twist is implanted, even
considering the simple evolution of the twist dictated by ohmic dissipation, 
an increase of the pulsed fraction in time is expected because, 
as time elapses, the bundle of current-carrying field lines shrinks and 
the heated area decreases. 
%In this respect we caveat that the blackbody fits of
%the first 4 observations show an increase in the blackbody radius. On the other
%hand, the statistics is such that we are not able to unambiguously confirm 
%the evolution of the emitting area,  are not conclusive in this respect. 
On the other hand, an increase in both pulsed fraction and emitting area, 
which is not excluded by the present data, would point toward a complete 
disassociation between the hot spot and the timing properties, with the 
latter perhaps of magnetospheric origin. The XRT statistics is not enough 
to unambiguously assess the evolution of the emitting area, nor even the 
presence of two distinct spectral components (BB and PL) detected by \xmm\ 
in 2006 and 2007, although there are several hints suggesting that also 
in the spectra observed during October 2008 two components are needed. 

% and the size of the former dictated by another form of heat release at the surface. 
%On the other hand, the measured increasing pulsed fraction as the emitting area  to a 
%different origin for the pulsed fraction which is also increasing 
%toward a complete disassociation among hot spot and pulsed fraction, with the
%latter perhaps of magnetospheric origin and the size of the former 
%dictated by another form of heat release at the surface. 

Overall, the spectral properties of \src\ are also very similar to those of other
transient magnetars, with an evident decay of the flux and a
softening of the emitted photons (at least in the 1--10 keV range) as the
outburst evolves in time. The spectral softening is apparent while using 
both the BB and the PL spectral models. Referring, for concreteness, to the 
power-law model, the photon
index switches from $\sim 2$ at the outburst onset to $\sim 4$ after $\sim 10$
d. A softening is expected if the spectrum originates in a globally twisted
magnetosphere by resonant up-scattering. As the twist angle decreases the charge
density decreases making scattering less efficient. This implies that the
high-energy tail filled up by up-scattered thermal photons becomes less and less
populated, producing a softening of the spectrum. This has been verified by
detailed Monte Carlo simulations in the case of a globally twisted dipolar field
(e.g. \citealt{nobili08}). Although no detailed model has been presented as yet,
it seems likely that the same picture holds if a localized twist is considered 
instead.  As it was mentioned earlier, the twist angle may still increase for a while   
following the onset of the outburst, which would be consistent with the negative 
$\ddot{P}$, while its spatial extent decreases \citep{belo09}. This latter effect 
may compensate for the first one and be the main effect responsible for  the spectral 
evolution:  in particular, if the region in which photons can be efficiently 
scattered by the currents is reduced, a flux softening and decrease in luminosity 
are  expected.

%Although this aspect of the bursts/outburst activity of magnetars
%seems rather understood, and well explained by the current models,
%there are two aspects which still miss a clear picture: 1) the physics
%driving the onset of the pulsed radio emission observed in connection
%to two AXPs' outbursts (as e.g. for the 2007 outburst of \src), and 2)
%the extremely different timescales of the outburst decay, which
%depending on the source can range from some weeks to years.

\section*{acknowledgements}
This research is based on observations with the NASA/UK/ASI \swift\ mission.
We thank the \swift\ duty scientists and science planners for making these
observations possible. The Italian authors acknowledge the partial support from
ASI (ASI/INAF contracts I/088/06/0, I/011/07/0, AAE~TH-058, AAE~DA-044, and AAE~DA-006). 
PE thanks the Osio Sotto city council for support with a G.~Petrocchi fellowship. SZ acknowledges support from STFC. NR is supported by a Ram\'on~y~Cajal fellowship. DG acknowledges the CNES for financial support.

\bibliographystyle{mn2e}
%\bibliography{biblio}

\bsp

\label{lastpage}

\end{document}